\pdfoutput=1
\documentclass[10pt, a4paper, conference, compsocconf]{IEEEtran}

\usepackage{makeidx}  
\usepackage{upgreek}

\usepackage{url}

\usepackage{amsmath}

\usepackage[T1]{fontenc}

\usepackage{array}

\usepackage{subfig}
\usepackage{multirow}
\usepackage{array}
\usepackage{amssymb}
\usepackage{float}

\usepackage[numbers]{natbib}

\usepackage{enumitem}

 \usepackage{amssymb}
 \usepackage{pifont}
\usepackage{textcomp}

\usepackage{upgreek}

\usepackage{url}

\usepackage{amsmath}

\usepackage[T1]{fontenc}

\usepackage{array}
\usepackage{graphicx}
\graphicspath{{./pdf/}{./jpeg/}{./images/}}
   \DeclareGraphicsExtensions{.pdf,.jpeg,.jpg,.png}
   
\usepackage{eurosym}

\begin{document}

\title{Towards the Leveraging of Data Deduplication to Break the Disk Acquisition Speed Limit}






\author{\IEEEauthorblockN{
Hannah Wolahan\IEEEauthorrefmark{4}, 
Claudio Chico Lorenzo\IEEEauthorrefmark{2}, 
Elias Bou-Harb\IEEEauthorrefmark{3}, 
Mark Scanlon\IEEEauthorrefmark{1}}
\IEEEauthorblockA{\\ \IEEEauthorrefmark{1}\IEEEauthorrefmark{4}School of Computer Science, University College Dublin, Ireland\\
\IEEEauthorrefmark{2}IES. Puerto de la Cruz, Spain\\
\IEEEauthorrefmark{3}Cyber Threat Intelligence Laboratory, Florida Atlantic University, USA\\\\
Email: \IEEEauthorrefmark{1}mark.scanlon@ucd.ie,
\IEEEauthorrefmark{4}hannah.wolahan@ucdconnect.ie,
\IEEEauthorrefmark{2}claudiochicolorenzo@gmail.com, \IEEEauthorrefmark{3}ebouharb@fau.edu}}

\maketitle

\begin{abstract}
Digital forensic evidence acquisition speed is traditionally limited by two main factors: the read speed of the storage device being investigated, i.e., the read speed of the disk, memory, remote storage, mobile device, etc.), and the write speed of the system used for storing the acquired data. Digital forensic investigators can somewhat mitigate the latter issue through the use of high-speed storage options, such as networked RAID storage, in the controlled environment of the forensic laboratory. However, traditionally, little can be done to improve the acquisition speed past its physical read speed from the target device itself. The protracted time taken for data acquisition wastes digital forensic experts' time, contributes to digital forensic investigation backlogs worldwide, and delays pertinent information from potentially influencing the direction of an investigation. In a remote acquisition scenario, a third contributing factor can also become a detriment to the overall acquisition time -- typically the Internet upload speed of the acquisition system. This paper explores an alternative to the traditional evidence acquisition model through the leveraging of a forensic data deduplication system. The advantages that a deduplicated approach can provide over the current digital forensic evidence acquisition process are outlined and some preliminary results of a prototype implementation are discussed.
\end{abstract}


\section{Introduction}
\label{intro}



One of the most significant issues facing the digital forensic communities around the world is attempting to deal with the sheer volume of cases requiring specialist analysis \citep{lillis2016challenges, scanlon2016deduplication}. A significant case backlog is commonplace in law enforcement agencies across the globe with legal proceedings often being heard with potentially pertinent evidence sitting in evidence storage lockers remaining unanalysed. This backlog is commonly in the order of 6-18 months, but can reach the order of years in some jurisdictions \citep{hitchcock2016tiered}. This delay in digital evidence processing can have a detrimental effect on the justice system \citep{casey2009investigation}. A large number of factors can contribute to this backlog, and a number of these are outlined below:

\begin{itemize}
\item Expertise Availability -- Law enforcement agencies have a limited amount of trained personnel to process the digital evidence in a forensically-sound, court-admissible manner.
\item Variety of Devices and Data Types -- The sheer diversity of devices types, storage media and remote sources of data greatly hinder digital investigative progress. With the advent of cloud-based storage of much of consumer level data, the traditionally simple device identification step of the digital forensic process has now become significantly more complex.
\item Protracted Acquisition and Analysis Time -- The acquisition and analysis phase are extremely time-heavy necessities of all digital forensics investigations. Acquiring evidence from physical or remote storage is limited by the read or upload speeds respectively.
\end{itemize}

While the first two factors can only be effectively tackled through training, education and expertise, this paper focuses on a technical solution capable of greatly expediting the last item in the above list. Through the leveraging of a deduplicated forensic data storage system, the acquisition and analysis phases of a typical investigation can be significantly expedited. Eliminating the unnecessary reacquisition and analysis of previously processed data can greatly expedite the current forensic process. The current digital forensic process typically starts with seizure of equipment from a suspect. This equipment is brought to an evidence store where it awaits further processing by expert personnel. Once the case reaches the top of the queue, the storage devices (hard drives, solid state drives, memory cards, external storage devices, etc.) are acquired using industry standard tools, e.g., EnCase or FTK Imager. This image (with a file size matching the original device) is then available for analysis on the forensic workstation or stored on a shared network storage device in the forensic laboratory.

%
%
%
%

\subsection{Contribution of this Work}
\label{contribution}
The contribution of this work can be summarised as:

\begin{itemize}[noitemsep,topsep=0pt]
\item The design of a data deduplication system capable of the secure, verifiable acquisition of digital storage media is outlined.
\item A method for the reconstruction of an entire disk image from a deduplicated data store.
\item The results of an evidence acquisition of numerous hard drive images to a proof-of-concept implementation of the deduplication system are presented and evaluated. 
\end{itemize}

\section{Related Work}
\label{related}

\subsection{Expediting the Digital Forensic Process}
\begin{figure*}
\centering
\includegraphics[width=\textwidth, trim=2cm 21cm 0cm 0cm, clip=true]{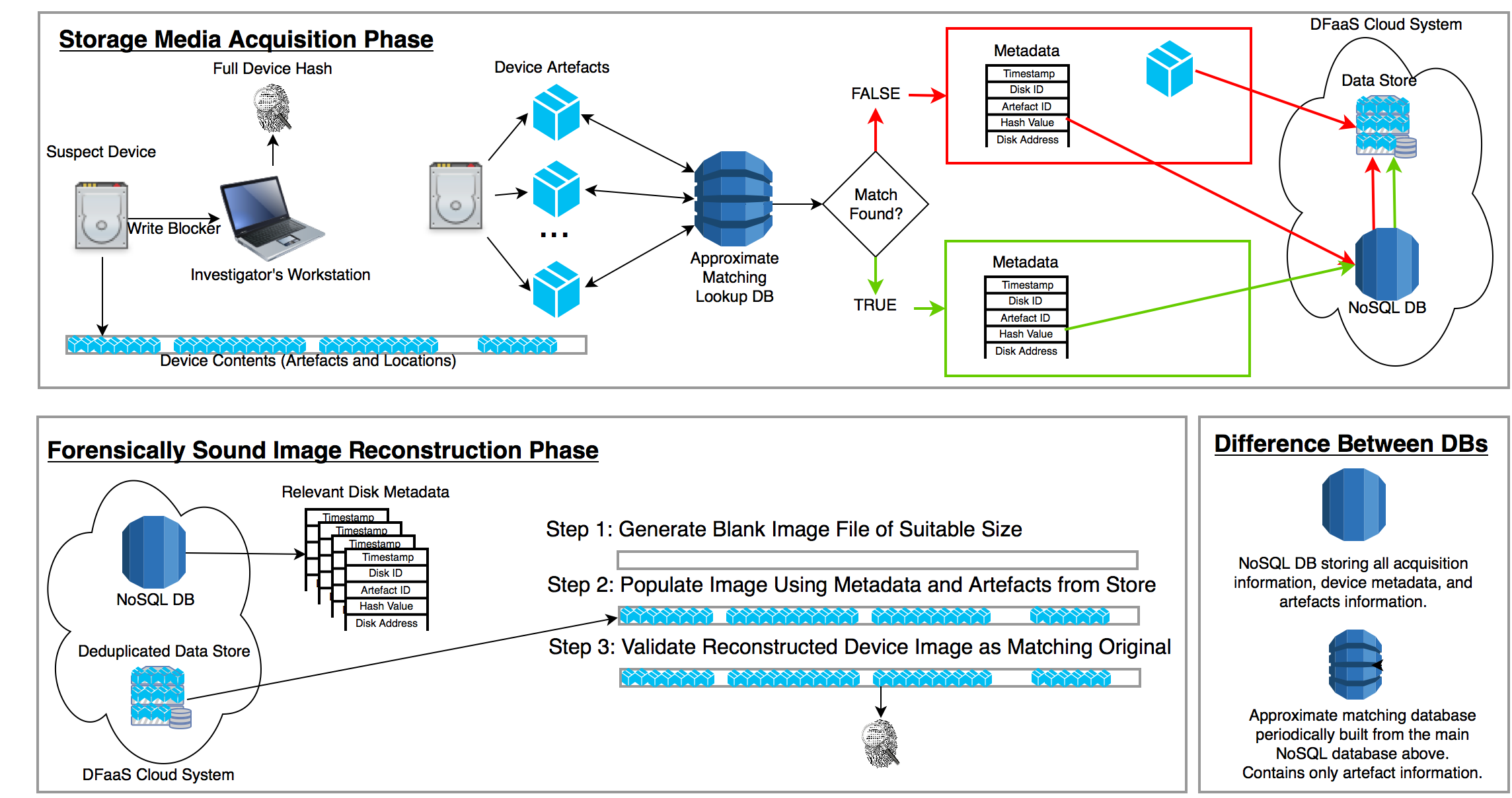}
\caption{Disk Image Acquisition Phase}
\label{fig:acquisition}
\end{figure*}
Current approaches to expedite the digital forensic investigative process focus on a range of aspects and techniques to speed up the analysis and contraband identification process. Most techniques to date broadly fall into the following approaches:

\begin{itemize}
\item Triage -- Digital forensic field triage involves providing minimal training to front-line law enforcement in the identification, handling and seizure of digital forensic evidence. An extension to this methodology is to provide remote access to suspect devices to digital forensic experts to enable time sensitive analysis and investigation. Fast detection of pertinent evidence can prove crucial to the timely progression of a number of investigation \citep{penrose2015fast}. \cite{rogers2006computer} proposed a formalised process model for the conducting of computer forensics field triage -- pushing some of the traditional forensic laboratory investigative steps to the field. \cite{hitchcock2016tiered} proposed a tiered methodology to enable based digital forensic triage capabilities by non-digital evidence specialists in order to facilitate the betting spending of expert time.
\item Remote Acquisition -- Many national police forces have a limited group of trained digital forensic evidence handling specialists \citep{hitchcock2016tiered}. Combining the deployment of digital forensic first responders with a remote evidence acquisition tool, such as that described by \cite{scanlon2010online}, would facilitate a more efficient utilisation of trained forensic experts. Uploading disk images from the field would facilitate forensic analysts to stay in the laboratory processing case work as opposed to wasting time out in the field.
\item Automation -- Expert experience and knowledge is predominantly used in conjunction with standard, and often basic, digital forensic tools to conduct the majority of digital investigations. While automation is an obvious improvement over the typical approach, it is not without its own challenges. \cite{james2013challenges} outline a number a challenges with automation in the digital forensic process including, with many centring around automated systems having a lack of knowledge capability of case specific information.
\end{itemize}

\subsection{Data Deduplication}
A deduplicated evidence storage system, such as those described by \citet{watkins2009teleporter} and \citet{scanlon2016deduplication}, can facilitate expedited evidence acquisition. Each unique file encountered need only be stored, indexed, analysed, and annotated once on the system. Eliminating non-pertinent, benign files during the acquisition phase of the investigation would immensely reduce the acquisition time (e.g., operating system, application, previously acquired non-incriminating files, etc.). This could greatly expedite pertinent information being available to the detectives working on the case as early as possible in the investigation. In order for any evidence to be court admissible, a forensically sound entire disk image would need to be reconstructible from the deduplicated data store, improving upon the system proposed by \citet{watkins2009teleporter}. Employing such a system would also facilitate a cloud-to-cloud based storage event monitoring of virtual systems as merely the changes of the virtual storage would need to be stored between each acquisition.

\begin{figure*}
\centering
\includegraphics[width=\textwidth, trim=2cm 0cm 17.4cm 21cm, clip=true]{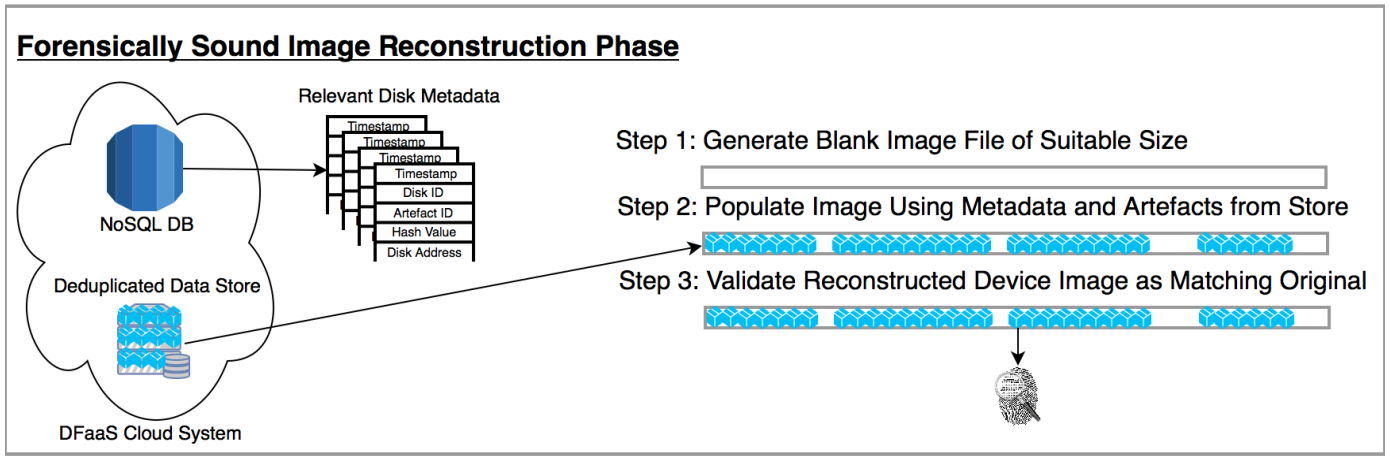}
\caption{Complete Disk Image Reconstruction Phase}
\label{fig:recon}
\end{figure*}

\subsection{Digital Forensics as a Service (DFaaS)}

Digital Forensics as a Service (DFaaS) is a cloud-based, modernised extension of current digital forensic investigative methods. While still very much in its infancy, some progress has been made in its development. In 2010, the Netherlands Forensic Institute (NFI) implemented a DFaaS solution in order to combat the volume of backlogged cases in the country \citep{van2014digital}. \cite{van2014digital} describe the advantages of the current system including efficient resource management, enabling detectives to directly query the data, improving the turnaround time between forming a hypothesis in an investigation its confirmation based on the evidence, and facilitating easier collaboration between detectives working on the same case through annotation and shared knowledge.

While the aforementioned DFaaS system is a significant step in the right direction, many improvements to the current model could greatly expedite and improve upon the current process. This includes improving the functionality available to the case detectives, improving its current indexing capabilities and on-the-fly identification of incriminating evidence during the acquisition process \citep{van2014digital}.

\section{System Design}
\label{arch}

\subsection{Acquisition}

Figure~\ref{fig:acquisition} shows an overview of the disk image acquisition phase of the proposed system. The acquisition phase follows the traditional model initially requiring the crime scene processing and seizure of pertinent equipment. Once the identification and seizure phases are complete, there are two usage scenarios for employing the proposed data deduplication system. The first scenario involves the traditional process: bring the devices back to a forensic laboratory and imaging the storage media from there. Most forensic laboratories would use a network attached storage system (NAS) for storing complete disk images. In this LAN-based scenario, the data would instead be directed to the deduplicated system. The second scenario is a new paradigm made more feasible through the use of a deduplicated forensic model: a remote evidence acquisition from the field or local forensic laboratory back to a central evidence storage and processing system, such as a national law enforcement DFaaS system. In this scenario, the digital evidence is gathered and transmitted over the Internet in a secure and verifiable manner.

Irrespective of which usage scenario is necessary, the actual acquisition process is similar. Once the devices are seized and ready for imaging, they are connected to a forensic workstation using a write-blocker. The client side application initially requires the identification and hashing of all storage device artefacts (partitions, files, slackspace, etc.) and compares each of these with the database of all previously acquired files. This comparison process can also be greatly expedited through the deployment of an approximate matching database. This approximate matching database can be frequently updated and generated on the server-side as files are added to the system. If a match is found, the associated metadata for this acquisition is recorded in the central database including the filename, creation and modification dates, the disk address, its fragmentation, and associated acquisition data such as the disk ID, case ID, investigator ID, etc. This metadata is linked to the deduplicated artefact stored on the server-side. If no such match is found, the same metadata is recorded alongside a transmission of a copy of the artefact itself to the server-side file repository.

In practice, local hashing, local metadata collection, comparison against the central repository, file uploading and indexing can all happen in a largely multi-threaded simultaneous fashion. Each file that is found to already exist on the server does not need to be reacquired effectively eliminating the transmission of this file (over the LAN or Internet) to the repository. However, in this model, the bottleneck of the disk-read speed is still present as each artefact must be hashed in order to determine its duplicate status. Fuzzy hashing techniques could be integrated into the system, which would greatly improve the hashing speed, while sacrificing a small amount of accuracy. As such, the overall throughput during the acquisition speed of the data from the disk can be greater than the read speed of the disk.

\subsection{Disk Image Reconstruction}

When it comes to the reconstruction of an entire disk image from the deduplicated data storage, it is a matter of gathering all of the metadata associated with a specific acquisition and populating the disk image file artefact by artefact, as can be seen in Figure~\ref{fig:recon}. A full-size blank disk image is first created as a staging area. Subsequently, each artefact is retrieved from the deduplicated data store and added into the disk image at the corresponding offset. Using this process, a complete disk image can be reconstructed with all of the corresponding artefacts found during the acquisition step. This reconstructed image would be verifiable as a true copy of the original through entire disk image hashing, e.g., to verify the court admissibility of the reconstructed evidence.

\section{Preliminary Experimentation and Results}
\label{results}

\subsection{Experimental Setup}

For the purposes of proving the viability of the proposed data deduplication system, a prototype client-server system was developed in Python using \texttt{pytsk} (python bindings for The Sleuth Kit). This section outlines some preliminary results of the acquisition phase using the system. \texttt{pytsk} facilities the processing of numerous file systems in python including NTFS, FAT12,
FAT16, FAT32, exFAT, UFS1, UFS1b, UFS2, Ext2, Ext3, Ext4, SWAP, RAW, ISO9660, HFS and YAFFS2, but for the purposes of the experimentation presented in this paper, NTFS was used.

Four base virtual machines were created with separate virtual hard disks in place for each of the 64-bit operating systems Windows 7, Windows Server 2012, Windows 8 and Windows 10. The experiments below were conducted over the Internet (simulating the aforementioned remote acquisition scenario).

\subsection{Initial Image Acquisition}

\begin{figure}
\centering
\includegraphics[width=0.5\textwidth]{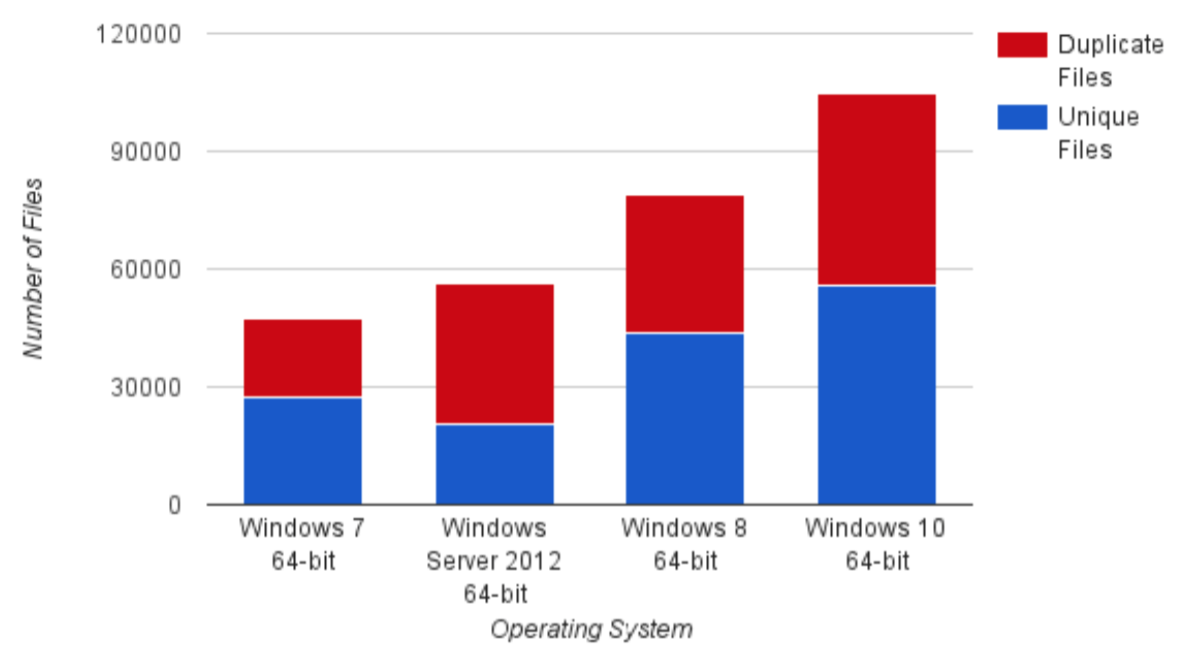}
\caption{Order of acquisitions to the deduplicated forensic storage system from L-R}
\label{fig:osfiles}
\end{figure}

Figure~\ref{fig:osfiles} shows the order of acquisition of the base images from left to right into a bare prototype, i.e., at the start of the Windows 7 acquisition, there was no files in the system. The level of duplication is portrayed in red, with the number of files identified during the Windows 7 duplication being those contained on the single disk, e.g., one file, \texttt{prnca00z.inf\_loc}, was found to occur 89 times on the system. Subsequently, the base disk images relating to the other operating systems were added to the system. The runtime for these initial operating system acquisitions are highlighted in Table~1. As might be expected, the overall runtime of these acquisitions is relatively slow compared to what one might anticipate from the traditional alternative due to the hashing and comparison steps and a lack of substantial artefact elimination.

\begin{figure}
\centering
\includegraphics[width=0.5\textwidth, trim=0cm 0cm 0cm 2cm, clip=true]{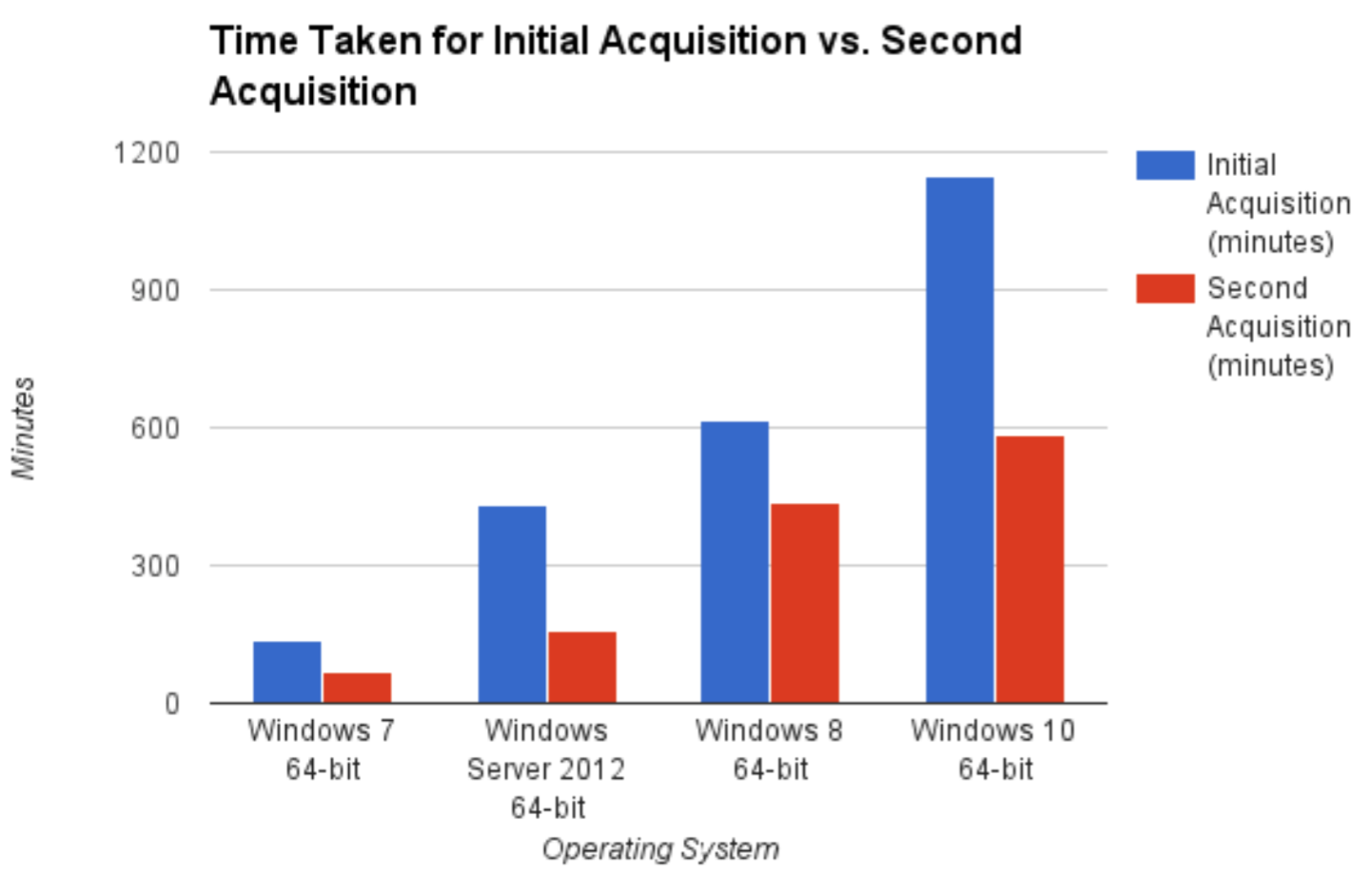}
\caption{Comparison of Initial Acquisition vs. Second Acquisition Time per bare Operating System Install}
\label{fig:firstsecond}
\end{figure}

Figure~\ref{fig:firstsecond} shows the improvement of a second acquisition of these base images with 100\% duplication. The runtime is greatly improved, however there is significant room for further improvement as the processing and evidence elimination phases used for testing were running single threaded. The precise time measurements for this test are also outlined in Table~1.

\begin{table*}[]
\centering
\label{tab:initialsecond}
\begin{tabular}{|l|c|c|}
\hline
\multicolumn{1}{|c|}{\textbf{Operating System}} & \textbf{\begin{tabular}[c]{@{}c@{}}Initial Acquisition\\ (Minutes)\end{tabular}} & \textbf{\begin{tabular}[c]{@{}c@{}}Second Acquisition\\ (Minutes)\end{tabular}} \\ \hline
Windows 7 64-bit                                & 139.14                                                                           & 66.87                                                                           \\ \hline
Windows Server 2012 64-bit                      & 429.42                                                                           & 158.35                                                                          \\ \hline
Windows 8 64-bit                                & 615.29                                                                           & 439.44                                                                          \\ \hline
Windows 10 64-bit                               & 1148.35                                                                          & 583.69                                                                          \\ \hline
\end{tabular}
\caption{Initial acquisition time for each disk image compared against its reacquisition}
\end{table*}

\subsection{Detection of Minor Modifications}

\begin{figure}
\centering
\includegraphics[width=0.3\textwidth]{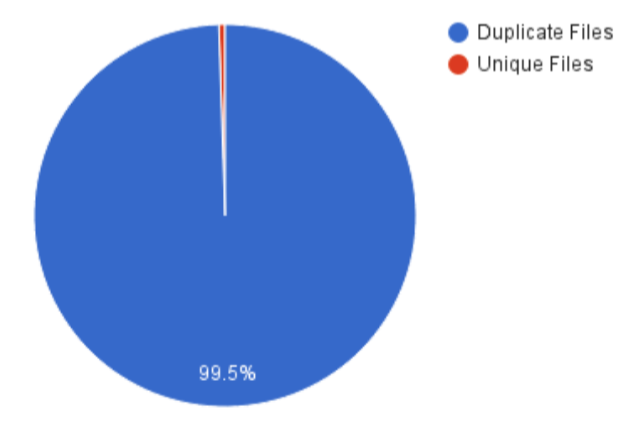}
\caption{Measurement of Modified Files between Boot Cycles of a Windows 10 Install}
\label{fig:piechart}
\end{figure}

For the purposes of this experiment, the Windows 10 system was booted up and used for regular Internet usage for a period of 30 minutes. In that time, a total of 0.5\% of the files (550) on the system were modified and identified when the deduplication system reacquired the data. For reference, this acquisition was added to a database already containing Windows 10 acquisitions. As a result, 99.5\% of the files on the disk were found to be duplicates in the system and hence, not reacquired.

\section{Concluding Remarks}
\label{conclusion}
While the initial acquisition to a bare system takes longer than what would be expected from the traditional alternative (as would be expected for each new operating system added to any deduplicated storage, i.e., maximum artefact uniqueness), subsequent acquisitions of the same machine or additional machines sharing many common files are greatly improved. As more and more acquisitions are fed into the system, the acquisition and analysis time required for each new device analysed would correspondingly decrease, i.e., higher duplicate rates would be achieved. Such a technique could be used for the monitoring of virtual machines or cloud instances whereby the resolution of the captures could be much higher than any of the alternatives available. More intelligent deduplicated evidence data storage and analysis techniques can help eliminate duplicated processing and duplicated expert analysis of previously content. Once a file has been analysed and determined to be incriminating, an alert can be generated during the acquisition phase to notify the investigator that illegal content has been discovered at the earliest stage possible in the investigation.

\subsection{Future Work}
\label{future}
The system outlined as part of this paper is in its infancy and there is much work to be done to improve the system and increase its functionality, performance and viability as a true replacement for the traditional digital forensic approach. A number of these avenues for research and development are outlined below:
\begin{itemize}
\item Improvements to the processing capabilities of the prototype system in terms of multi-threading, fuzzy hashing, automated evidence classification, etc.

\item Combining expert contributed analysis of acquired files facilitating automated blacklisting or whitelisting of suspect devices. Functionality such as this can greatly help towards the elimination of duplicated efforts in the analysis of content.

\item Using the proposed deduplication system can also greatly expedite the acquisition of digital evidence from hash-based file-synchronisation services, such as BitTorrent Sync or Syncthing \citep{scanlon2015network, quinn2015forensic}. Given that these tools rely on frequent hashing to determine when a synchronisation is required, those same hashes can be used for artefact acquisition and elimination purposes.
\end{itemize}

\newpage

{\footnotesize \bibliographystyle{plainnat}
\bibliography{./bibliography}}

\begin{thebibliography}{12}
\providecommand{\natexlab}[1]{#1}
\providecommand{\url}[1]{\texttt{#1}}
\expandafter\ifx\csname urlstyle\endcsname\relax
  \providecommand{\doi}[1]{doi: #1}\else
  \providecommand{\doi}{doi: \begingroup \urlstyle{rm}\Url}\fi

\bibitem[Casey et~al.(2009)Casey, Ferraro, and Nguyen]{casey2009investigation}
Eoghan Casey, Monique Ferraro, and Lam Nguyen.
\newblock Investigation delayed is justice denied: proposals for expediting
  forensic examinations of digital evidence.
\newblock \emph{Journal of forensic sciences}, 54\penalty0 (6):\penalty0
  1353--1364, 2009.

\bibitem[Hitchcock et~al.(2016)Hitchcock, Le-Khac, and
  Scanlon]{hitchcock2016tiered}
Ben Hitchcock, Nhien-An Le-Khac, and Mark Scanlon.
\newblock Tiered forensic methodology model for digital field triage by
  non-digital evidence specialists.
\newblock \emph{Digital Investigation}, 16\penalty0 (S1):\penalty0 75--85, 03
  2016.
\newblock ISSN 1742-2876.
\newblock Proceedings of the Third Annual DFRWS Europe.

\bibitem[James and Gladyshev(2013)]{james2013challenges}
Joshua~I James and Pavel Gladyshev.
\newblock Challenges with automation in digital forensic investigations.
\newblock \emph{arXiv preprint arXiv:1303.4498}, 2013.

\bibitem[Lillis et~al.(2016)Lillis, Becker, O'Sullivan, and
  Scanlon]{lillis2016challenges}
David Lillis, Brett Becker, Tadhg O'Sullivan, and Mark Scanlon.
\newblock {Current Challenges and Future Research Areas for Digital Forensic
  Investigation}.
\newblock 05 2016.

\bibitem[Penrose et~al.(2015)Penrose, Buchanan, and
  Macfarlane]{penrose2015fast}
Philip Penrose, William~J Buchanan, and Richard Macfarlane.
\newblock Fast contraband detection in large capacity disk drives.
\newblock \emph{Digital Investigation}, 12:\penalty0 S22--S29, 2015.

\bibitem[Quinn et~al.(2015)Quinn, Scanlon, Farina, and
  Kechadi]{quinn2015forensic}
Conor Quinn, Mark Scanlon, Jason Farina, and M-Tahar Kechadi.
\newblock Forensic analysis and remote evidence recovery from syncthing: An
  open source decentralised file synchronisation utility.
\newblock In \emph{Digital Forensics and Cyber Crime}, pages 85--99. Springer,
  2015.

\bibitem[Rogers et~al.(2006)Rogers, Goldman, Mislan, Wedge, and
  Debrota]{rogers2006computer}
Marcus~K Rogers, James Goldman, Rick Mislan, Timothy Wedge, and Steve Debrota.
\newblock Computer forensics field triage process model.
\newblock In \emph{Proceedings of the conference on Digital Forensics, Security
  and Law}, page~27. Association of Digital Forensics, Security and Law, 2006.

\bibitem[Scanlon(2016)]{scanlon2016deduplication}
Mark Scanlon.
\newblock {Battling the Digital Forensic Backlog through Data Deduplication}.
\newblock In \emph{{Proceedings of the 6th IEEE International Conference on
  Innovative Computing Technologies (INTECH 2016)}}, Dublin, Ireland, 08 2016.
  IEEE.

\bibitem[Scanlon and Kechadi(2009)]{scanlon2010online}
Mark Scanlon and M-Tahar Kechadi.
\newblock Online acquisition of digital forensic evidence.
\newblock In \emph{Proceedings of International Conference on Digital Forensics
  and Cyber Crime (ICDF2C 2009)}, pages 122--131. Springer, Albany, New York,
  USA, September 2009.

\bibitem[Scanlon et~al.(2015)Scanlon, Farina, and Kechadi]{scanlon2015network}
Mark Scanlon, Jason Farina, and M-Tahar Kechadi.
\newblock Network investigation methodology for bittorrent sync: A peer-to-peer
  based file synchronisation service.
\newblock \emph{Computers and Security}, 54:\penalty0 27 -- 43, 10 2015.
\newblock ISSN 0167-4048.
\newblock \doi{http://dx.doi.org/10.1016/j.cose.2015.05.003}.
\newblock URL
  \url{http://www.sciencedirect.com/science/article/pii/S016740481500067X}.

\bibitem[Van~Baar et~al.(2014)Van~Baar, van Beek, and van Eijk]{van2014digital}
RB~Van~Baar, HMA van Beek, and EJ~van Eijk.
\newblock Digital forensics as a service: A game changer.
\newblock \emph{Digital Investigation}, 11:\penalty0 S54--S62, 2014.

\bibitem[Watkins et~al.(2009)Watkins, McWhorte, Long, and
  Hill]{watkins2009teleporter}
Kathryn Watkins, Mike McWhorte, Jeff Long, and Bill Hill.
\newblock {Teleporter: An Analytically and Forensically Sound Duplicate
  Transfer System}.
\newblock \emph{Digital Investigation}, 6:\penalty0 S43--S47, 2009.

\end{thebibliography}


\end{document}